\documentclass[aps,pra,twocolumn,amsmath,amssymb,nofootinbib,showpacs,superscriptaddress]{revtex4-1}

\usepackage[english]{babel}
\usepackage{latexsym}
\usepackage{graphics}
\usepackage{graphicx}
\usepackage{epsfig}
\usepackage{color}
\usepackage{bm}
\usepackage{amsmath}
\usepackage{amssymb}
\usepackage{amsthm}
\usepackage{dcolumn}
\usepackage{bm}
\usepackage{float}
\usepackage{hyperref}
\usepackage{epstopdf}
\usepackage{cleveref}
\usepackage[svgnames]{xcolor}
\usepackage{enumerate}
\hypersetup{hidelinks,colorlinks=true,allcolors=DarkBlue}

\newcommand{\ket}[1]{\left|{#1}\right\rangle} 
\newcommand{\bra}[1]{\left\langle{#1}\right|} 

\usepackage{lineno}

\begin{document}
	
	\preprint{APS/123-QED}
	
	\title{Estimating the precision for quantum process tomography}
	
	\author{E.O. Kiktenko}
	\affiliation{Russian Quantum Center, Skolkovo, Moscow 143025, Russia}
	\affiliation{Moscow Institute of Physics and Technology, Dolgoprudny, Moscow Region 141700, Russia} 
	\affiliation{Department of Mathematical Methods for Quantum Technologies, Steklov Mathematical Institute of Russian Academy of Sciences, Moscow 119991, Russia}
	
	\author{D.N. Kublikova}
	\affiliation{Russian Quantum Center, Skolkovo, Moscow 143025, Russia}
	\affiliation{Moscow Institute of Physics and Technology, Dolgoprudny, Moscow Region 141700, Russia} 
	\affiliation{Schaffhausen Institute of Technology, Schaffhausen 8200, Switzerland}
	\affiliation{National University of Singapore, Singapore 117546, Singapore}
	
	\author{A.K. Fedorov}
	\affiliation{Russian Quantum Center, Skolkovo, Moscow 143025, Russia}
	\affiliation{Moscow Institute of Physics and Technology, Dolgoprudny, Moscow Region 141700, Russia} 
	
	\date{\today}
	\begin{abstract}
		Quantum tomography is a widely applicable tool for complete characterization of quantum states and processes.
		In the present work, we develop a method for precision-guaranteed quantum process tomography.
		With the use of the Choi--Jamiołkowski isomorphism, we generalize the recently suggested extended norm minimization estimator for the case of quantum processes.
		Our estimator is based on the Hilbert-Schmidt distance for quantum processes.
		Specifically, we discuss the application of our method for characterizing quantum gates of a superconducting quantum processor in the framework of the IBM Q Experience. 
	\end{abstract}
	\maketitle
	
	\section{Introduction}
	
	A task of the complete characterization of quantum states and processes is an essential part of quantum technologies~\cite{NielsenChuang}.
	Quantum tomography allows complete characterization and assessment of a state via carrying out measurements over multiple copies of the state in various bases~\cite{Lvovsky2009}.
	An important part of tomographic protocols is evaluating the accuracy of the output result.
	This problem has been widely studied~\cite{Bogdanov2009,Blume-Kohout2012,Renner2012,Flammia2012,Flammia2011,Silva2011,Sugiyama2013,Faist2016,Wang2019}. 
	A useful technique for solving this task, which is known as precision-guaranteed quantum state tomography, has been proposed in Ref.~\cite{Sugiyama2013}.
	This approach has a number of important advantages over other existing approaches.
	First, it can guarantee that the state estimated by a quantum tomography protocol is arbitrarily close to the true state with a high probability if the number of measurements is sufficient.
	Moreover, the bounds on a distance measure between the reconstructed density matrix and true density matrix for given confidence level (CL)  can be extracted from experiments.
	As the distance measure between the reconstructed density and true density matrices, Hilbert-Schmidt distance, trace distance, and infidelity have been considered. 
	
	Although the precision-guaranteed quantum state tomography is helpful for evaluating the precision of the quantum states preparation, 
	quantum information processing systems also require an approach for evaluating the quality of operations with quantum states.
	One of the ways to solve this problem is to use the randomized benchmarking technique~\cite{Knill2008}.
	Much more detailed information can be obtained via quantum process tomography~\cite{Poyatos1997,White2004,Bantysh2018,Leonhardt1997,D'Ariano,Chuang,D'Ariano2001,Mohseni2008}, which is a natural extension of quantum state tomography.
	Quantum process tomography allows predicting the effect of the quantum processes on arbitrary input states.
	This technique has been used in a number of experiments and, in particular, for characterization of a controlled-NOT (\textsf{CNOT}) gate~\cite{White2004} in the linear-optics experiment.
	We note that two-qubit entangling gates, such as \textsf{CNOT}, are fundamental elements in archetypal gate-based quantum computers.
	
	\begin{figure}
		\includegraphics[width=0.85\linewidth]{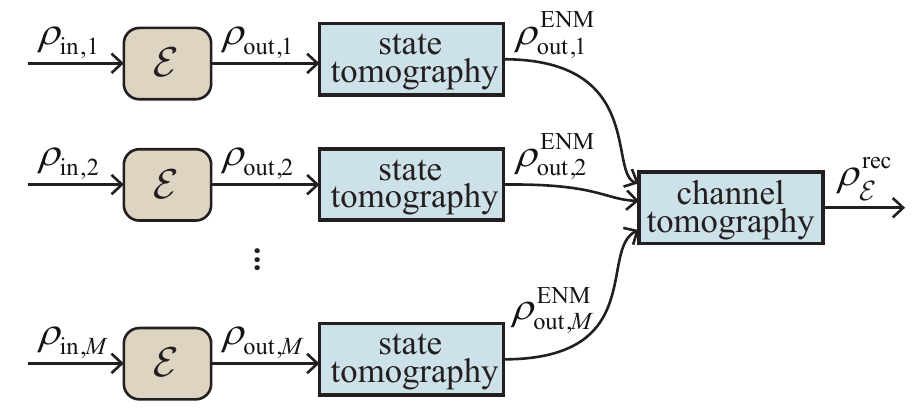}
		\caption{The basic scheme of the precision-guaranteed process tomography.}
		\label{fig:genscheme}
	\end{figure}
	
	The question of precision is also of great importance for quantum process tomography protocols. 
	Data processing in quantum process tomography experiments can be simplified via the Choi--Jamiołkowski isomorphism~\cite{Jamiolkowski1972,Choi1975}.
	The Choi--Jamiołkowski isomorphism establishes a correspondence between linear completely positive (CP) maps $\mathcal{E}$ 
	from operators on the Hilbert space $\mathcal{H}$ to space $\mathcal{K}$ and positive semidefinite operators on the Hilbert space $\mathcal{H}\otimes\mathcal{K}$. 
	The problem of the experimental characterization of quantum processes could be then reduced to the estimation of a specific quantum state in the Hilbert space of a higher dimension, which is known as a Choi state.
	The task of complete characterization of quantum processes is of special interest in the context of the characterization of noisy intermediate-scale quantum devices.
	Indeed, one can think of using precision-guaranteed quantum-process tomography techniques for estimating the quality of quantum gates. 
	
	In this work, we develop a method for quantum process tomography with guaranteed precision. 
	For this task, we use the Choi--Jamiołkowski isomorphism for the generalization of the approach for estimating the precision of quantum-state tomography for the case of quantum processes (see Fig.~\ref{fig:genscheme}). 
	We develop our estimators on the basis of the Hilbert-Schmidt distance. 
	The suggested approach allows obtaining not only an estimate of an unknown quantum state or process, but also estimating a distance to target states or processes. 
	We apply our technique for experimental characterization of quantum gates in a superconducting quantum processor.
	
	Our work is organized as follows.
	In Sec.~\ref{sec:pgQST}, we consider the precision-guaranteed quantum-state tomography protocol.
	In Sec.~\ref{sec:pgQPT}, we describe the suggested generalization of the precision-guaranteed technique for the case of quantum processes. 
	In Sec.~\ref{sec:experiment}, we use the proposed approach for characterizing quantum gates in a quantum processor. 
	We summarize our results in Sec.~\ref{sec:conclusion}.
	
	\section{Precision-guaranteed quantum state tomography}\label{sec:pgQST}
	
	Here we first briefly review the precision-guaranteed quantum-state tomography protocol~\cite{Sugiyama2013}.
	We consider a $d$-dimensional Hilbert space $\mathcal{H}$ and a quantum state $\rho$ given by a positive semi-definite unit-trace linear Hermitian operator acting on $\mathcal{H}$. 
	Consider a set of measurements ${\Pi}=\{ {\Pi}^{(j)}\}_{j=1}^{J}$, 
	where each ${\Pi}^{(j)}=\{\Pi_m^{(j)}\}_{m=1}^{M^{(j)}}$ is a positive operator-valued measure (POVM).
	The operators $\Pi_m^{(j)}$, called effects, satisfy the natural requirements: 
	$\Pi_m^{(j)}\geq0, \sum_m\Pi_m^{(j)}={I}$, where ${I}$ is the identity operator in $\mathcal{H}$.
	
	Let us introduce a vector consisting of orthonormal operators ${\bf \lambda}=(\lambda_1,\ldots,\lambda_{d^2-1})$, for which one has ${\rm tr}\lambda_i=0$ and ${\rm tr}(\lambda_i\lambda_j)=2\delta_{ij}$, where $\delta_{ij}$ is the Kronecker symbol.
	Then any density matrix $\rho$ can be parametrized in the following way:
	\begin{equation}
	\rho = \rho({\bf s})=\frac{1}{d}{I}+\frac{1}{2}({\bf s},{\bf\lambda}),
	\end{equation}
	where ${\bf s}=(s_1,\ldots,s_{d^2-1})$ is a vector of real parameters and $(\cdot,\cdot)$ stands for the standard dot product between vectors.
	One can also write down the decomposition of POVM effects as follows:
	\begin{equation}
	\Pi_{m}^{(j)}=a_{m, 0}^{(j)} \mathbf{I}+(\mathbf{a}_{m}^{(j)}, {\bf \lambda}),
	\end{equation}
	where $a_{m, 0}^{(j)}$ are real, and $\mathbf{a}_{m}^{(j)}$ is $d^2-1$ real vector.
	
	\begin{figure}[h]
		\includegraphics[width=0.45\linewidth]{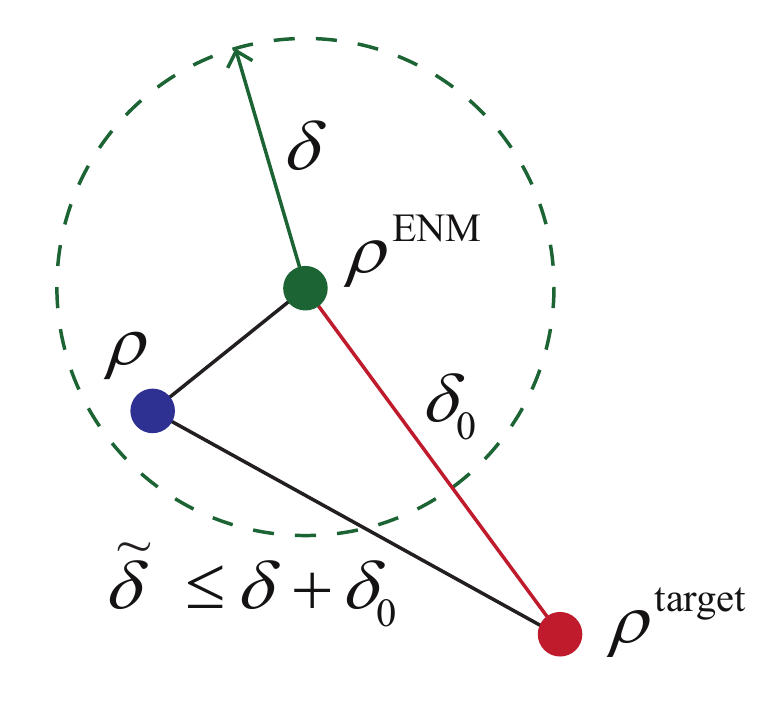}
		\vskip -4mm
		\caption{The idea behind the precision-guaranteed quantum state tomography protocol.}
		\label{fig:est}
	\end{figure}
	
	We then consider a problem of calculating an estimate of the state $\rho$ from the measurement result obtained by measuring ${\Pi}$.
	Let each ${\Pi}^{(j)}$ was measured $n^{(j)}$ times and each of the effects $\Pi^{(j)}_i$ was obtained $n_{i}^{(j)}$ times ($\sum_{i=1}^{M^{(j)}} n_{i}^{(j)}=n^{(j)}$).
	We can construct a linear least squares (LLS) estimator given by the following expression:
	\begin{equation}
	\begin{split}
	&\rho^{\rm LLS}:=\rho({\bf s}^{\rm LLS}), \quad {\bf s}^{\rm LLS}=A_L({\bf f}-{\bf a}_0), \\
	&A_L := (A^TA)^{-1}A^T,
	\end{split}
	\end{equation}
	where $(i,(j,m))$th element of $A$ is given by $a^{(j)}_{m,i}$, $f_{(j,m)}=n^{(j)}_m/n^{(j)}$, $a_{0,(j,m)}=a^{(j)}_{m,0}$.
	By definition $\rho^{\rm LLS}$ satisfies a requirement ${\rm tr}\rho^{\rm LLS}=1$. 
	However, some of the eigenvalues may appear to be negative, which corresponds to a not physically plausible state.
	
	In order to satisfy the physical conditions, we introduce an extended norm minimization (ENM) and obtain the physical estimate in the following form:
	\begin{equation}
	\rho^{\rm ENM} :=  {\arg\min}_{\rho^{\prime} \geq 0, {\rm tr} \rho^{\prime}=1}
	\left[
	\Delta^{\mathrm{HS}}\left(\rho_{N}^{\mathrm{LLS}}, \rho^{\prime}\right)
	\right]^2,
	\end{equation}
	where
	\begin{equation}
	\Delta^{\mathrm{HS}}\left(\rho, \rho^{\prime}\right) :=2^{-1/2}\left[{\rm tr}\left(\rho-\rho^{\prime}\right)^{2}\right]^{1 / 2}
	\end{equation}
	is the Hilbert-Schmidt distance.
	$\rho^{\rm ENM}$ can be obtained by removing terms with negative eigenvalues from the spectral decomposition of $\rho^{\rm LLS}$ followed by renormalization to obtain a unit trace.
	
	In Ref.~\cite{Sugiyama2013} it is proven the following result for $\rho^{\rm ENM}$:
	\begin{equation}\label{eq:PGST}
	\Pr\left[\Delta^{\mathrm{HS}}\left(\rho^{\mathrm{ENM}}, \rho\right) \leq \delta\right] \geq \mathrm{CL},
	\end{equation}
	where
	\begin{equation}
	\mathrm{CL} :=1-2 \sum_{\alpha=1}^{d^{2}-1} \exp \left[-\frac{8}{\left(d^{2}-1\right) c_{\alpha}} \delta^{2} N\right]
	\end{equation}
	is CL defined via
	\begin{eqnarray}
	c_{\alpha} :=&&\sum_{j=1}^{J} r^{(j)} \left(\max _{m}\left[A_{L}^{-1}\right]_{\alpha,(j, m)}-\min _{m}\left[A_{L}^{-1}\right]_{\alpha,(j, m)}\right)^2, \nonumber \\ 
	&&r^{(j)} :=\frac{N}{n^{(j)}}, \quad N := \sum_j n^{(j)}.
	\end{eqnarray}
	
	Eq.~(\ref{eq:PGST}) can be interpreted as follows: 
	With probability at least ${\rm CL}$ our estimate $\rho^{\mathrm{ENM}}$ is no further than $\delta$ from real state $\rho$ (in terms of Hilbert-Schmidt distance).
	We note that Eq.~(\ref{eq:PGST}) also allows us to estimate a distance between the real state $\rho$ and some target state $\rho^{\rm target}$.
	Using the triangle inequality, one obtains the following relation [see also Fig.~\ref{fig:est}]:
	\begin{equation}
	\Delta^{\rm HS}(\rho,\rho^{\rm target}) \leq 
	\Delta^{\rm HS}(\rho^{\rm ENM},\rho^{\rm target})+\Delta^{\rm HS}(\rho^{\rm ENM},\rho).
	\end{equation}
	We note that $\Delta^{\rm HS}(\rho^{\rm ENM},\rho^{\rm target})=:\delta_0$ can be easily calculated.
	Thus, finally we obtain the following expression: 
	\begin{equation}
	\Pr\left[\Delta^{\mathrm{HS}}\left(\rho, \rho^{\rm target}\right) \leq \widetilde{\delta}\right] \geq \mathrm{CL}, \quad \widetilde{\delta} := \delta+\delta_0.
	\end{equation}
	This estimate can provide information about the quality of the preparation process for a given state and help to perform its calibration.
	The next stage is to extend this approach for the case of quantum processes.

	\section{Quantum process tomography with guaranteed precision}\label{sec:pgQPT}
	
	For estimating the quality of operations with quantum states, the discussed above precision-guaranteed technique should be extended for the case of quantum processes.
	Let us consider a completely positive trace preserving (CPTP) map $\mathcal{E}$ transferring linear operators acting on some Hilbert space $\mathcal{H}_{\rm in}$ to linear operators acting on Hilbert space $\mathcal{H}_{\rm out}$.
	
	Our technique allows obtaining a bound $\widetilde{\Delta}$ on the Hilbert-Schmidt distance between the target Choi state $\rho_\mathcal{E}^{\rm target}$ and true Choi state given by
	\begin{equation}
	\rho_{\mathcal{E}}=\frac{1}{d_{\rm in}} \sum_{n, m}|n\rangle\langle m|\otimes\sigma_{n,m}, \quad \sigma_{n,m}=\mathcal{E}[|n\rangle\langle m|]
	\end{equation}
	of an unknown process $\mathcal{E}[\cdot]$ for a given confidence level ${\rm CL}$ such that
	\begin{equation}\label{eq:channel_prec_tom}
	\Pr\left[\Delta^{\rm HS}(\rho_\mathcal{E},\rho_\mathcal{E}^{\rm target})\leq\widetilde{\Delta}\right]\geq{\rm CL}.
	\end{equation}
	
	First, we consider a problem of obtaining a point estimate $\rho_\mathcal{E}^{\rm rec}$ of an unknown Choi state $\rho_\mathcal{E}$ by performing tomography of a set of output states 
	$\{\rho_{{\rm out},i}=\mathcal{E}[\rho_{{\rm in},i}]\}_{i=1}^M$ obtained from known set of input states $\{\rho_{{\rm in},i}\}_{i=1}^M$ (see also Fig.~\ref{fig:genscheme}).
	Here we assume that the set $\{\rho_{{\rm in},i}\}_{i=1}^M$ forms an (over)complete basis in the space of linear operators acting in the Hilbert space of input states.
	
	In order to obtain $\rho_{\mathcal{E}}^{\rm rec}$, let us rewrite the Choi in the following form:
	\begin{equation}
	\rho_{\mathcal{E}}=\frac{1}{d_{\rm in}} \sum_{n, m}|n\rangle\langle m|\otimes\sum_kC_{n,m}^k\rho_{{\rm out},k},
	\end{equation}
	where coefficients $C_{n,m}^k$ arrive from the decompositions:
	\begin{equation}
	\ket{n}\bra{m}=\sum_{k}C_{n,m}^k\rho_{{\rm in},k}.
	\end{equation}
	Let $\rho_{{\rm out},i}^{\rm ENM}$ be a result of the tomographic reconstruction of $\rho_{{\rm out},i}$ with the use of the precision-guaranteed state tomography protocol (which is described above).
	We assume that each $\rho_{{\rm out},i}^{\rm ENM}$ is obtained with same measurements and number of samples, so we can introduce a pair (${\rm CL}, \delta$) such that 
	\begin{equation}
	\Delta^{\rm HS}(\rho_{{\rm out},k},\rho_{{\rm out},k}^{\rm ENM})=2^{-1/2}\sqrt{ {\rm tr}\left[\chi_k^2\right] }\leq \delta 
	\end{equation}
	with the probability at least ${\rm CL}$, where $\chi_k:=\rho_{{\rm out},k}-\rho_{{\rm out},k}^{\rm ENM}$.
	Next, let us introduce a state
	\begin{equation}
	\rho_{\mathcal{E}}^{\rm ENM}:=\frac{1}{d_{\rm in}} \sum_{n, m}|n\rangle\langle m|\otimes\sum_kC_{n,m}^k\rho_{{\rm out},k}^{\rm ENM}
	\end{equation}
	and consider its projection 
	\begin{equation}
	\rho_{\mathcal{E}}^{\rm rec} := \arg\min_{\rho'\in{\mathcal{S}}} \|\rho'-\rho_{\mathcal{E}}^{\rm ENM}\|_2
	\end{equation}
	on the space $\mathcal{S}$ of physical Choi states, which are semi-positive and whose partial trace over the first subsystem gives a maximally mixed state.
	In order to obtain the projection, one can employ techniques proposed in Ref.~\cite{Knee2018}.
	
	In order to estimate an upper bound on the Hilbert-Schmidt distance $\Delta^{\rm HS}(\rho_\mathcal{E},\rho_\mathcal{E}^{\rm rec})$
	we use the following inequality:
	\begin{equation}
	\begin{split}
	\Delta^{\rm HS}(\rho_\mathcal{E},\rho_\mathcal{E}^{\rm rec})\leq \Delta^{\rm HS}(\rho_\mathcal{E},\rho_\mathcal{E}^{\rm ENM})= \\ 
	\frac{1}{d_{\rm in}\sqrt{2}}\left\{
	{\rm tr}\left[
	\left(
	\sum_{n,m}\ket{n}\bra{m}\otimes\sum_k C_{n,m}^k \chi_k
	\right)^2
	\right]
	\right\}^{\frac{1}{2}},
	\end{split}
	\end{equation}
	
	One can see that 
	\begin{equation}
	\begin{split}
	{\rm tr}
	\left[
	\left(
	\sum_{n,m}\ket{n}\bra{m}\otimes\sum_k C_{n,m}^k \chi_k
	\right)^2
	\right]= \\
	\sum_{k,k'}{\rm tr}[\chi_k\chi_{k'}]\sum_{n,m}C_{n,m}^kC_{m,n}^{k'}
	\leq 2\delta^2\sum_{k,k'} \left| 
	\sum_{n,m}C_{n,m}^k \left(C_{n,m}^k\right)^*
	\right|,
		\end{split}
	\end{equation}
	in the view of the H\"older's inequality
	\begin{equation}
	{\rm tr}\left[
	\chi_k\chi_k'
	\right]\leq \sqrt{
		{\rm tr}\left[\chi_k^2\right]
		{\rm tr}\left[\chi_k^{'2}\right]
	} \leq 2\delta^2.
	\end{equation}
	
	As a results, we arrive at the following inequality:
	\begin{equation} \label{eq:Delta}
	\Delta^{\rm HS}(\rho_\mathcal{E},\rho_\mathcal{E}^{\rm rec})\leq
	\frac{\delta}{d_{\rm in}}\sqrt{
		\sum_{k,k'} \left| 
		\sum_{n,m}C_{n,m}^k\left(C_{n,m}^{k'}\right)^*
		\right|	
	}=:\Delta,
	\end{equation}
	and the statement about precision of reconstructing $\rho_{\mathcal{E}}$ reads
	\begin{equation}
	\Pr\left[\Delta^{\rm HS}(\rho_{\mathcal{E}},\rho_{\mathcal{E}}^{\rm rec})\leq \Delta \right] \geq {\rm CL}.
	\end{equation}
	In order to obtain the desired estimate~\eqref{eq:channel_prec_tom} one just need to set
	\begin{equation}
	\widetilde{\Delta} := \Delta+\Delta^{\rm HS}(\rho_{\mathcal{E}}^{\rm target},\rho_{\mathcal{E}}^{\rm rec}).
	\end{equation}
	
	Finally, we note that for given values of $\delta$ and ${\rm CL}$, determined by the state-tomography protocol, the accuracy of the process-tomography is determined by coefficients $\{C_{n,m}^k\}$.
	Thus, the considered approach allows finding an optimal set of probe states whose tomography after passing them through the channel gives the best reconstructing accuracy.
	
	It is also important to note that in our consideration we rely on H\"older's inequality, that is possible due to the definition of the Hilbert-Schmidt distance.
	One should be careful with the Hilbert-Schmidt distance since two completely orthogonal mixed states $\rho_1$ and $\rho_2$ with ${\rm tr}(\rho_1,\rho_2)=0$ can have $\Delta^{\rm HS}(\rho_1,\rho_2)\sim d^{-1/2}\rightarrow 0$ with $d\rightarrow \infty$.
	However, this problem does not emerge in low-dimensional Hilbert spaces which are the main case study of our work.
	
	\section{Characterization of quantum gates}\label{sec:experiment}
	
\begin{figure*}
		\includegraphics[width=1\linewidth]{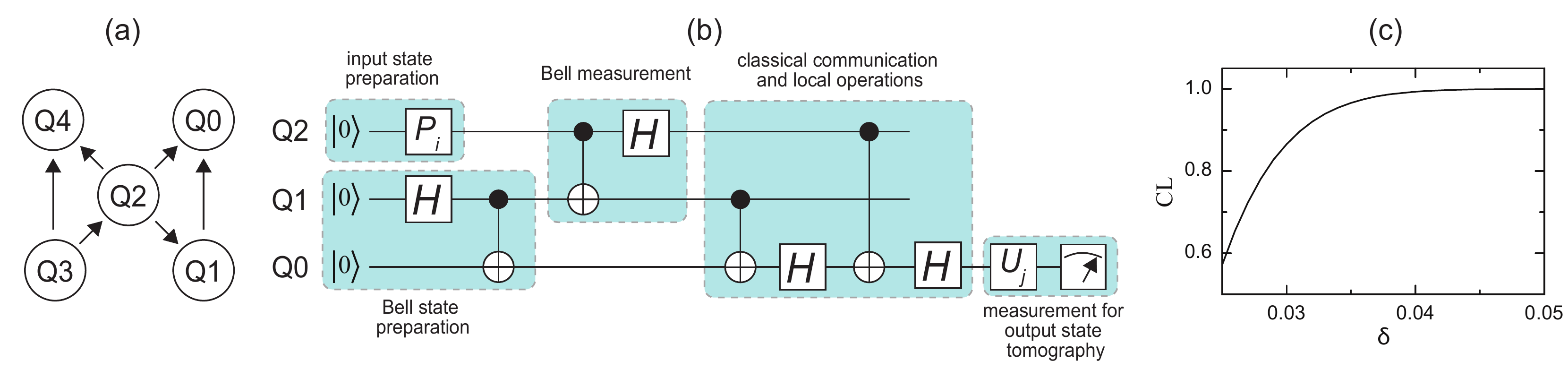}
		\caption{In (a) the scheme of connection in the IBM Q5.1 processor is presented. 
			The arrows between physical qubits correspond to the ability to perform immediate \textsf{CNOT} gates (arrow is directed from the control qubit to the target one). 
			In (b) the general scheme of the teleportation circuit tomography is shown. 
			Here an arbitrary quantum state prepared on Q2 is transferred and then measured on Q0.
			In (c) the confidence level CL is presented as a function of the Hilbert-Schmidt distance upper bound $\delta$ in the employed precision-guaranteed state tomography.
		}
		\label{fig:telep}
	\end{figure*}

	The developed technique is of particular interest for quantum computing platforms, where any computation is imperfect due to the presence of decoherence and experimental imperfections.
	We employ the considered approach to tomography of the quantum teleportation protocol, which is simulated on IBM Q5.1 superconducting quantum processor (see Fig.~\ref{fig:telep}).
	
	The considered three-qubit circuit simulates a single-qubit teleportation protocol via a shared entangled Bell state~\cite{NielsenChuang}.
	The circuit simulates all the steps of the protocol including the Bell state construction, Bell measurement on Alice's side, classical communication from Alice to Bob, and local operations on Bob's side.
	In the ideal case, the teleportation protocol corresponds to the identical channel, 
	which transfers the state space of the physical qubit Q2 to the state space of the physical qubit Q0 with the corresponding Choi state $\rho_{\rm Id}=\ket{{\Phi}^+}\bra{{\Phi}^+}$ with $\ket{\Phi^+}=2^{-1/2}(\ket{00}+\ket{11})$. 
	In reality, the resulting map is given by some different Choi state $\rho_{\rm telep}$ mainly due to imperfections of two-qubit controlled not (\textsf{CNOT}) gates.
	We note that in our consideration we focus on two-qubit gate errors and then assume that errors of qubits initializations in zero states and errors of all single qubit gates are negligible.
	Surely, they exist in realistic quantum computing devices, however, there are several technique for the suppression of their role. 
	
	\begin{table}
		\begin{tabular}{c|c|c|c}
			$i$ & $\rho_{{\rm out},i}^{\rm ENM}$ & $\delta_i$ & $\widetilde{\delta}_i$ \\ \hline
			1 &
			$\begin{bmatrix}
			0.969 &  -0.038-0.027{\rm i} \\
			-0.038+0.027{\rm i}  & 0.031
			\end{bmatrix}$
			& 0.03 & 0.06\\
			2 &
			$\begin{bmatrix}
			0.420 &   0.009+0.348{\rm i} \\
			0.009-0.348{\rm i} & 0.580
			\end{bmatrix}$
			& 0.03 & 0.15\\
			3 &
			$\begin{bmatrix}
			0.430 &  -0.38-0.193{\rm i} \\
			-0.380+0.193{\rm i}  & 0.570 
			\end{bmatrix}$
			& 0.03 & 0.11 \\
			4 &
			$\begin{bmatrix}
			0.419 &  0.235-0.288{\rm i}  \\
			0.235+0.288{\rm i}  & 0.581
			\end{bmatrix}$
			& 0.03 & 0.20 \\
		\end{tabular}
		\caption{Experimental results of the precision guaranteed state tomography of output states for ${\rm CL}=0.87$.}
		\label{tab:states}
	\end{table}
	
	\begin{table*}
		\begin{tabular}{c|c|c}
			$\rho_{\rm tel}^{\rm rec}$ & $\Delta$ & $\widetilde{\Delta}$ \\ \hline
			$\begin{bmatrix}
			0.484 &   -0.019-0.013{\rm i} &  -0.004 &    0.372+0.058{\rm i}  \\
			-0.019+0.013{\rm i} &  0.016   & -0.020 &   0.003-0.002{\rm i} \\
			-0.003-0.002{\rm i} &    -0.002 &   0.075 &    -0.024-0.027{\rm i} \\
			0.397+0.058{\rm i} &  0.003+0.002{\rm i} & -0.024+0.027{\rm i}  & 0.425 					
			\end{bmatrix}$ & 0.03 & 0.18
		\end{tabular}
		\caption{Experimental results of the precision guaranteed process tomography based on the results from the Table~\ref{tab:states} (${\rm CL}=0.87$).}
		\label{tab:channel}
	\end{table*}
	
	In order to perform the process tomography protocol, we employ a set of four pure input states
	\begin{equation}
		\rho_{{\rm in},i}=\ket{\psi_i}\bra{\psi_i},\quad
		\ket{\psi_i}=P_i\ket{0}, 
		\quad i
		\in
		\{1,2,3,4\}
	\end{equation}
	which are prepared by using the corresponding set of unitary operators
	\begin{equation}
	\begin{split}
	& P_1 = {I}, \,\, 
	P_2=R_x(\theta_0),\,\, 
	P_3=R_z\left(\frac{4\pi}{3}\right)R_x(\theta_0),\,\,  \\
	& P_4=R_z\left(\frac{2\pi}{3}\right)R_y(\theta_0),\,\, 
	\theta_0=\arccos\left(-\frac{1}{3}\right).
	\end{split}
	\end{equation}
	Here and after $R_x(\cdot)$, $R_y(\cdot)$, $R_z(\cdot)$ stand for standard rotation operations around the corresponding axis of the Bloch sphere~\cite{NielsenChuang}.
	One can see that the set $\{\ket{\psi}_i\}_{i=1}^4$ forms a regular tetrahedron in the Bloch sphere.
	One can check that this choice of the input states provides a minimum of $\Delta$ for given $\delta$ and $d_{\rm in}=2$ in Eq.~\eqref{eq:Delta}.
	
	Unlike the case of single-qubit gates, imperfections of the final computational basis measurement are taken into account, and after the calibration procedure we obtained its positive-operator valued measure (POVM) in the following form:
	\begin{equation}
	\Pi^{(3)} = \left\{
	\begin{bmatrix}
	0.972 & 0 \\ 0 &  0.093
	\end{bmatrix},
	\begin{bmatrix}
	0.028 & 0 \\ 0 & 0.907
	\end{bmatrix}
	\right\}.
	\end{equation}
	With the use of preceding unitary operators we obtained an informationally-complete set of POVMs
	${\Pi}=\{\Pi^{(1)}, \Pi^{(2)}, \Pi^{(3)}\}$,
	with	$\Pi^{(i)}_j = U_i^\dagger\Pi_j^{(3)}U_i$ and
	\begin{equation}
	U_1=R_y\left(\frac{\pi}{2}\right),\quad
	U_2=R_x\left(\frac{\pi}{2}\right),\quad
	U_3={I}.
	\end{equation}
	
	The results of the precision-guaranteed state tomography protocol of output states $\{\rho_{{\rm out},i}\}_{i=1}^4$ is presented in Table~\ref{tab:states}.
	The measurement statistics is obtained with $n^{(i)}=8192$ circuit runs for each configuration of input state preparation and output state measurement (thus, in total we have  $N_{\rm tot}=3\cdot4\cdot8192=98304$ runs of the circuits).
	The upper bound on the Hilbert-Schmidt distance $\delta_i$ between point estimates $\rho_{{\rm out},i}^{\rm ENM}$ and true output states $\rho_{{\rm out},i}$ for a given CL appeared to be the same for all inputs since we use the same measurement setup. 
	The dependence of CL on $\delta:=\delta_i$ is presented in Fig.~\ref{fig:telep}(c).
	For our further consideration we fix a point $\delta=0.03$ and ${\rm CL}=0.87$.
	In the Table~\ref{tab:states} we also show an upper bound on distances $\widetilde{\delta}_i$ to ideal (target) output states equal to the input ones $\rho_{{\rm out},i}^{\rm target}=\rho_{{\rm in},i}$.
	
	The experimental results of the whole precision-guaranteed process tomography are summarized in Table~\ref{tab:channel}.
	Due to the choice of the input states set, the upper bound $\Delta$ on the Hilbert-Schmidt distance for the given ${\rm CL}$ between real Choi state and the reconstructed one appeared to be equal to $\delta$.
	Using a distance to the ideal (target) Choi--Jamiolkowiski state $\rho_{\rm Id}$ we can formulate the final results of the precision-guaranteed approach: 
	\begin{equation}
	\begin{aligned}
	&\Pr[\Delta^{\rm HS}(\rho_{\rm tel},\rho_{\rm tel}^{\rm rec})\leq 0.03]\geq 0.87,	\\
	&\Pr[\Delta^{\rm HS}(\rho_{\rm tel},\rho_{\rm Id})\leq 0.18]\geq 0.87.
	\end{aligned}
	\end{equation}
	
	One can see that the uncertainty of the reconstructed Choi--Jamiolkowiski state due to the final statistics is much lower than the distance to the target state.
	Thus, one can conclude that imperfections of $\textsf{CNOT}$ gates have the significant impact on the resulting state.
	Our technique gives a qualitative analysis of this impact. 
	
	\section{Conclusion and outlook}\label{sec:conclusion}
	
	Here we summarize the main results of our work.
	We have demonstrated a method for precision-guaranteed quantum process tomography.
	We have applied this technique for characterizing quantum gates in the superconducting quantum processor in the framework of the IBM Q Experience.
	The obtained results can be used for precision-guaranteed quantum process tomography in quantum systems of various nature, including superconducting circuits, atomic arrays, and photonic devices.
	
	We also note that a straightforward approach to quantum process tomography is to probe the process with a set of states whose density operators form a spanning set in the space of all operators over a particular Hilbert space~\cite{Poyatos1997}. 
	However, this approach requires a large set of difficult-to-prepare probe states, and is consequently restricted to to low-dimensional systems. 
	One of the possible solutions is to use coherent quantum states only for characterizing quantum processes~\cite{Lobino2008,Keshari2011,Anis2012,Fedorov2015,Leonhardt1997}.
	Therefore, an important task is to extend the suggested method for evaluating the precision of quantum processes for the case of  coherent states as probes.
	Moreover, the promising direction of the research is employing the precision guaranteed tomography in self-calibrating protocols~\cite{Branczyk2012}.
	Another interesting point is to use this technique for investigating possible advantages of machine learning approaches for tomography of quantum states and processes~\cite{Troyer2018,Tiunov2019}.
	
%	\medskip
	
	\section*{Acknowledgments}
	
	We acknowledge the use of the IBM Q Experience for this work.
	The views expressed are those of the authors and do not reflect the official policy or position of IBM or the IBM Q Experience team.
	The authors thank A.A. Karazeev, A.E. Ulanov, E.S. Tiunov, V.V. Tiunova, and A.I. Lvovsky for fruitful discussions and useful comments. 
	The present work was supported by the grant of the President of the Russian Federation (project MK- 923.2019.2).


\begin{thebibliography}{99}
		
		\bibitem{NielsenChuang}
		M.A. Nielsen and I.L. Chuang,	
		{\it Quantum Computation and Quantum Information}
		(Cambridge University Press, 2000).
		
		\bibitem{Lvovsky2009}
		A.I. Lvovsky and M.G. Raymer, 
		Continuous-variable optical quantum-state tomography,
		{\href{https://doi.org/10.1103/RevModPhys.81.299}{Rev. Mod. Phys. {\bf 81}, 299 (2009)}}.
		
		\bibitem{Bogdanov2009}
		Yu.I. Bogdanov,
		Unified statistical method for reconstructing quantum states by purification,
		\href{https://doi.org/10.1134/S106377610906003X}{J. Exp. Theor. Phys. {\bf 108} 928 (2009)}. 
		
		\bibitem{Blume-Kohout2012}
		R. Blume-Kohout, 
		Robust error bars for quantum tomography,
		{\href{https://arxiv.org/abs/1202.5270}{arXiv.org:1202.5270}}.
		
		\bibitem{Renner2012}
		M. Christandl and R. Renner, 
		Reliable quantum state tomography,
		{\href{https://doi.org/10.1103/PhysRevLett.109.120403}{Phys. Rev. Lett. {\bf 109}, 120403 (2012)}}.
		
		\bibitem{Flammia2012}
		S.T. Flammia, D. Gross, Y.-K. Liu, and J. Eisert, 
		Quantum tomography via compressed sensing: error bounds, sample complexity and efficient estimators,
		{\href{https://doi.org/10.1088/1367-2630/14/9/095022}{New J. Phys. {\bf 14}, 095022 (2012)}}.
		
		\bibitem{Flammia2011}
		S.T. Flammia and Y.-K. Liu, 
		Direct fidelity estimation from few Pauli measurements,
		{\href{https://doi.org/10.1103/PhysRevLett.106.230501}{Phys. Rev. Lett. {\bf 106}, 230501 (2011)}}.
		
		\bibitem{Silva2011}
		M.P. da Silva, O. Landon-Cardinal, and D. Poulin, 
		Practical characterization of quantum devices without tomography,
		{\href{http://dx.doi.org/10.1103/PhysRevLett.107.210404}{Phys. Rev. Lett. {\bf 107}, 210404 (2011)}}.
		
		\bibitem{Sugiyama2013} 
		T. Sugiyama, P.S. Turner, and M. Murao, 
		Precision-guaranteed quantum tomography,
		{\href{http://dx.doi.org/10.1103/PhysRevLett.111.160406}{Phys. Rev. Lett. {\bf 111}, 160406 (2013)}}.
		
		\bibitem{Faist2016}
		P. Faist and R. Renner,
		Practical and reliable error bars in quantum tomography,
		\href{https://doi.org/10.1103/PhysRevLett.117.010404}{Phys. Rev. Lett. {\bf 117}, 010404 (2016)}.
		
		\bibitem{Wang2019}
		J. Wang, S.B. Volkher, and R. Renner,
		Confidence polytopes in quantum state tomography,
		\href{https://doi.org/10.1103/PhysRevLett.122.190401}{Phys. Rev. Lett. {\bf 122}, 190401 (2019)}.
		
		\bibitem{Knill2008} 	
		E. Knill, D. Leibfried, R. Reichle, J. Reichle, R.B. Blakestad, J.D. Jost, C. Langer, R. Ozeri, S. Seidelin, and D.J. Wineland,
		Randomized benchmarking of quantum gates
		\href{https://doi.org/10.1103/PhysRevA.77.012307}{Phys. Rev. A {\bf 77}, 012307 (2008)}.	
		
		\bibitem{Poyatos1997}
		J.F. Poyatos, J.I. Cirac, and P. Zoller, 
		Complete characterization of a quantum process: The two-bit quantum gate,
		{\href{http://dx.doi.org/10.1103/PhysRevLett.78.390}{Phys. Rev. Lett. {\bf 78}, 390 (1997)}}.
		
		\bibitem{White2004}
		J.L. O'Brien, G.J. Pryde, A. Gilchrist, D.F.V. James, N.K. Langford, T.C. Ralph, and A.G. White,
		Quantum process tomography of a controlled-NOT gate,
		{\href{http://dx.doi.org/10.1103/PhysRevLett.93.080502}{Phys. Rev. Lett. {\bf 93}, 080502 (2004)}}.	
		
		\bibitem{Bantysh2018}
		B.I. Bantysh, D.V. Fastovets, and Yu.I. Bogdanov,
		High-fidelity quantum tomography with imperfect measurements,
		{\href{https://doi.org/10.1117/12.2522413}{Proc. SPIE {\bf 11022}, 110222N (2018)}}.
		
		\bibitem{D'Ariano}
		G.M. D'Ariano, M. De Laurentis, M.G.A. Paris, A. Porzio, and S. Solimeno,
		Quantum tomography as a tool for the characterization of optical devices,
		{\href{https://doi.org/10.1088/1464-4266/4/3/366}{J. Opt. B {\bf 4}, 127 (2002)}}.
		
		\bibitem{Chuang}
		I.L. Chuang and M.A. Nielsen,
		Prescription for experimental determination of the dynamics of a quantum black box,
		{\href{http://dx.doi.org/10.1088/1367-2630/17/4/043063}{J. Mod. Opt. {\bf 44}, 2455 (1997)}}.
		
		\bibitem{D'Ariano2001}
		G.M. D'Ariano and P. Lo Presti, 
		{\href{http://dx.doi.org/10.1103/PhysRevLett.86.4195}{Phys. Rev. Lett. {\bf 86}, 4195 (2001)}}.
		
		\bibitem{Mohseni2008}
		M. Mohseni, A.T. Rezakhani, and D.A. Lidar,
		Quantum-process tomography: Resource analysis of different strategies,
		{\href{http://dx.doi.org/10.1103/PhysRevLett.86.4195}{Phys. Rev. A {\bf 77}, 032322 (2008)}}.	
		
		\bibitem{Jamiolkowski1972}
		A. Jamiołkowski,
		Linear transformations which preserve trace and positive semidefiniteness of operators,
		\href{https://doi.org/10.1016/0034-4877(72)90011-0}{Rep. Math. Phys. {\bf 3}, 275 (1972)}.
		
		\bibitem{Choi1975}
		M-D. Choi,
		Completely positive linear maps on complex matrices,
		\href{https://doi.org/10.1016/0024-3795(75)90075-0}{Linear Algebra Appl. {\bf 10}, 285 (1975)}.
		
		\bibitem{Knee2018}
		G.C. Knee, E. Bolduc, J. Leach, and E.M. Gauger,
		Quantum process tomography via completely positive and trace-preserving projection,
		\href{https://doi.org/10.1103/PhysRevA.98.062336}{Phys. Rev. A {\bf 98}, 062336 (2018)}.
		
		\bibitem{Lobino2008}
		M. Lobino, D. Korystov, C. Kupchak, E. Figueroa, B.C. Sanders, and A.I. Lvovsky,
		Complete characterization of quantum-optical processes,
		{\href{http://dx.doi.org/10.1126/science.1162086}{Science {\bf 322}, 563 (2008)}}.
		
		\bibitem{Keshari2011}
		S. Rahimi-Keshari, A. Scherer, A. Mann, A.T. Rezakhani, A.I. Lvovsky, and B.C. Sanders, 
		Quantum process tomography with coherent states,
		{\href{http://dx.doi.org/10.1088/1367-2630/13/1/013006}{New J. Phys. {\bf 13}, 013006 (2011)}}.
		
		\bibitem{Anis2012}
		A. Anis and A.I. Lvovsky,  
		Maximum-likelihood coherent-state quantum process tomography
		{\href{http://dx.doi.org/10.1088/1367-2630/14/10/105021}{New J. Phys. {\bf 14}, 105021 (2012)}}.
		
		\bibitem{Fedorov2015}
		I.A. Fedorov, A.K. Fedorov, Y.V. Kurochkin, and A.I. Lvovsky,  
		Tomography of a multimode quantum black box,
		{\href{http://dx.doi.org/10.1088/1367-2630/17/4/043063}{New J. Phys. {\bf 17}, 043063 (2015)}}.
		
		\bibitem{Leonhardt1997}
		U. Leonhardt, 
		{\it Measuring the quantum state of light}, 
		(Cambridge University Press, Cambridge, 1997).
				
		\bibitem{Branczyk2012}
		A.M. Brańczyk, D.H. Mahler, L.A. Rozema, A. Darabi, A.M. Steinberg, and D.F.V. James,
		Self-calibrating quantum state tomography,
		\href{https://doi.org/10.1088/1367-2630/14/8/085003}{New J. Phys. {\bf 14}, 085003 (2012)}.
		
		\bibitem{Troyer2018}
		G. Torlai, G. Mazzola, J. Carrasquilla, M. Troyer, R. Melko, and G. Carleo,
		Neural-network quantum state tomography,
		{\href{http://dx.doi.org/10.1038/s41567-018-0048-5}{Nat. Phys. {\bf 14}, 447 (2017)}}.
		
		\bibitem{Tiunov2019}
		E.S. Tiunov, V.V. Tiunova, A.E. Ulanov, A.I. Lvovsky, and A.K. Fedorov,
		Experimental quantum homodyne tomography via machine learning,
		{\href{https://arxiv.org/abs/1907.06589}{arXiv.org:1907.06589}}.
		

	\end{thebibliography}
\end{document}